# Power System Transient Stability Analysis Using Truncated Taylor Expansion Systems

Bin Wang[1,2], *Member, IEEE*, Xin Xu[1], *Student Member, IEEE*, Kai Sun[1], *Senior Member, IEEE*
1. University of Tennessee, Knoxville, TN    2. Texas A&M University, College Station, TX

*Abstract*—Small signal analysis is a special case of analytical approaches using Taylor expansions of power system differential equations with the truncation performed at order one. The truncated Taylor expansions (TTEs) at higher orders can lead to better approaches for stability analysis by considering higher order nonlinearities, e.g. normal form, modal series and nonlinear modal decoupling. This paper presents fundamental studies on how accurate transient stability analysis results can be obtained from the TTE systems compared to that on the original system. The analytical investigation is conducted on single-machine-infinite-bus power systems. Conclusions are drawn from there and verified on two multi-machine power systems by extensive numerical simulations.

## I. INTRODUCTION

Modern power grids are operated with more and more stressed loading conditions, pushing them closer to their stability limits. Dynamic security assessment (DSA) plays an important role to guarantee the dynamic performance being within the limit under a list of credible contingencies, where the transient stability analysis is performed by either the time domain simulation, analytical approaches like direct methods or hybrid approaches. The transient stability analysis is to determine whether the post-disturbance initial condition locates inside the domain of attraction (DOA) of the stable equilibrium point (SEP) of the post-disturbance system. The time domain simulation is accurate but time-consuming, while the direct method is fast but less accurate. An accurate and fast transient stability analysis is always desired in practice and has been the goal of many continuous academic efforts made in the past several decades [1]. For faster time domain simulations, techniques like semi- analytical solutions [2][3] and high performance computers based parallel computing [4] are proposed. For more accurate analytical approaches, improved determination of the critical point and energy functions are also proposed [1][5].

Due to the simplicity in the derivation and calculation of polynomial functions, the truncated Taylor expansion (TTE) has already been adopted for power system stability analysis. The eigenanalysis, as a special case of the Taylor expansion (TE) truncated at order one, is used to study the power system small signal stability. Many nonlinear analyses, e.g. normal form [6], modal series method [7] and nonlinear modal decoupling [8], are proposed to study power system stability under large disturbances, where the reference [8] shows a possibility to use a 3$^{rd}$ order TTE for transient stability analysis. Note that all these analyses start from the TTEs at a certain order, but none of them showed to which order the resulting analysis will be accurate or inaccurate. What is the minimum order for TTE to study transient stability? Are the TTE based transient stability analysis conservative or optimistic? Is a high order TTE always necessary? The higher, the better? Answers to these questions are very important to any analytical approaches starting from the TTE, which will be the focus of this paper. This paper will aim at answering this question.

The remainder of the paper is organized as follows. The transient stability analyses using TTE systems at different orders are presented for single-machine-infinite-bus (SMIB) power systems in Section II and for multi-machine power systems in Section III. Conclusions are drawn in Section IV.

## II. TRANSIENT STABILITY ANALYSIS OF SMIB POWER SYSTEMS USING TTE SYSTEMS

Consider the swing equation of a general SMIB power system in classical model [9]. For simplicity, denote $\alpha = D/2H$ and $\beta = P_{max}\omega_s/2H$ and obtain (1).

$$\ddot{\delta} + \alpha\dot{\delta} + \beta(\sin\delta - \sin\delta_s) = 0 \qquad (1)$$

The SEP of (1) is $\delta = \delta_s < \pi/2$ and its two closest unstable equilibrium points (UEPs) around the SEP are $\delta_{u1} = \pi - \delta_s$ and $\delta_{u2} = -\pi - \delta_s$. Without loss of generality, assume the single machine is generating active power, i.e. $\delta_s > 0$. Then we have $|\delta_{u1}| < |\delta_{u2}|$. When applying the transient energy function (or equivalently the equal area criterion in the SMIB case) for stability analysis, the damping is usually ignored, i.e. $\alpha = 0$,

This work was supported by NSF grants ECCS-1553863 and ECCS-1610025

which simplifies the derivations a lot while introduces some conservativeness to the resulting stability analysis. Then, $\delta_{u1}$ is used to determine the critical energy, while $\delta_{u2}$ does not have any contribution here. Next will present how accurately the UEP $\delta_{u1}$ can be approximated from a TTE.

The TE of (1) at the SEP is (2), keeping only the first $n$ terms gives an approximate system in (3), called the $n$th order TTE system, or the TTE system at order $n$.

$$\ddot{\delta}+\alpha\dot{\delta}+\beta\sum_{k=1}^{\infty}\frac{\sin(\delta_s+k\pi/2)}{k!}(\delta-\delta_s)^k=0 \quad (2)$$

$$\ddot{\delta}+\alpha\dot{\delta}+\beta\sum_{k=1}^{n}\frac{\sin(\delta_s+k\pi/2)}{k!}(\delta-\delta_s)^k=0 \quad (3)$$

In the following, we rigorously derive the results for TTE systems with $n = 2$ and 3, and investigate cases with $n = 4$, 5,…, 9 by numerical studies. It is worth mentioning that although investigated by numerical approaches, the conclusions on these cases are valid since the only parameter in all expressions, i.e. $\delta_s$, is considered by the enumeration method with a high resolution. Also, note that the two cases with $n = 4$ and 5 do have closed-form solutions, but whose expressions are too complex to give a straightforward interpretation, so we simply investigate them numerically.

TABLE I    APPROXIMATION OF $\delta_{u1}$ BY TTE SYSTEMS AT ORDERS 2 & 3

| $n$ / order of TTE system | Approximation of $\delta_{u1}$ |
|---|---|
| ∞* | $\delta_{u1}=\pi-\delta_s$ |
| 2 | $\delta_{u1\_TE2}=\delta_s+\dfrac{2\cos\delta_s}{\sin\delta_s}$ |
| 3 | $\delta_{u1\_TE3}=\delta_s+\dfrac{\sqrt{9+15\cos^2\delta_s}-3\sin\delta_s}{2\cos\delta_s}$ |

* The TTE system with $n = \infty$ represents the original system in (1).

### A. Transient stability analysis using 2$^{nd}$ order TTE system

This subsection proves that the $\delta_{u1\_TE2}$ determined by the 2$^{nd}$ order TE system is always greater than $\delta_{u1}$, such that the stability analysis using 2$^{nd}$ order TTE system always gives optimistic results, i.e. an initial condition leading to a stable trajectory in the 2$^{nd}$ order TTE system may still cause instability in the original system.

**Claim 1**: $\delta_{u1\_TE2} > \delta_{u1}$ for any $\delta_s$ in $(0, \pi/2)$.

*Proof*: Define the difference between $\delta_{u1\_TE2}$ and $\delta_{u1}$ as $y$ and change the variable from $\delta_s$ to $x$ by $x = \cos\delta_s$. Then, claim 1 is equivalent to prove $y(x) > 0$ for any $x$ in $(0,1)$.

$$y(x)=2\arccos x+\frac{2x}{\sqrt{1-x^2}}-\pi \quad (4)$$

Note that $y(0) = 2\times\pi/2-\pi = 0$ and $y(1^-)=+\infty >0$ as shown in (5), so we only need $y'(x) < 0$ in (0,1), which is true in (6).

$$y(1^-)=\lim_{x\to 1^-}y(x)=\lim_{x\to 1^-}2\arccos x+\frac{2x}{\sqrt{1-x^2}}-\pi \quad (5)$$
$$=2\times 0+\infty-\pi=+\infty$$

$$y'(x)=\frac{2x^2}{(1-x^2)^{3/2}}>0 \quad (6)$$

### B. Transient stability analysis using 3$^{rd}$ order TTE system

This subsection proves that the $\delta_{u1\_TE3}$ determined by the 3$^{rd}$ order TTE is always smaller than $\delta_{u1}$, such that the stability analysis using 3$^{rd}$ order TTE system always gives conservative results, i.e. an initial condition leading to an stable trajectory in the 3$^{rd}$ order TTE system guarantees a stable trajectory in the original system.

**Claim 2**: $\delta_{u1\_TE3} > \delta_{u1}$ for any $\delta_s$ in $(0, \pi/2)$.

*Proof*: Define the difference between $\delta_{u1\_TE3}$ and $\delta_{u1}$ as $y$ and change the variable from $\delta_s$ to $x$ by $x = \cos\delta_s$. Then, claim 2 is equivalent to prove $y(x) < 0$ for any $x$ in $(0,1)$.

$$y(x)=2\arccos x+\frac{\sqrt{9+15x^2}-3\sqrt{1-x^2}}{2x}-\pi \quad (7)$$

Note that $y(0^+) = 0$ as shown in (8) and $y(1) = 2\times 0+\sqrt{6}-\pi \approx -0.6921<0$, so we only need to show $y'(x)<0$ in $(0,1)$, which is equivalent to show (10). For the right inequality in (10), when $x\geq 1/2$, it holds since the left side is non-positive; when $x<1/2$, it also holds since (11).

$$y(0^+)=\lim_{x\to 0^+}2\arccos x-\pi+\frac{\sqrt{9+15x^2}-3\sqrt{1-x^2}}{2x}$$
$$=2\times\pi/2-\pi+\lim_{x\to 0^+}\frac{\sqrt{9+15x^2}-3\sqrt{1-x^2}}{2x} \quad (8)$$
$$=\lim_{x\to 0^+}\frac{15x/\sqrt{9+15x^2}+x/\sqrt{1-x^2}}{2}=0$$

$$y'(x)=\frac{1/2}{x^2\sqrt{1-x^2}}\left(1-4x^2-\sqrt{\frac{9(1-x^2)}{1+5x^2/3}}\right) \quad (9)$$

$$1-4x^2-\sqrt{\frac{9(1-x^2)}{1+5x^2/3}}<0 \Leftrightarrow (1-4x^2)\sqrt{\frac{1+5x^2/3}{1-x^2}}<3 \quad (10)$$

$$(1-4x^2)\sqrt{\frac{1+5x^2/3}{1-x^2}}<(1-4\times 0^2)\sqrt{\frac{1+5\times(0.5)^2/3}{1-(0.5)^2}} \quad (11)$$
$$\approx 1.3744<3$$

### C. Transient stability analysis using high order TTE systems

The equation to determine $\delta_{u1\_TEn}$ is shown in (12), as did for 2$^{nd}$ and 3$^{rd}$ order cases, the smallest root that is greater than $\delta_s$ is selected as $\delta_{u1\_TEn}$ to approximate $\delta_{u1}$. Note that $\delta_{u1\_TEn}$ is a function of the only parameter $\delta_s$.

$$\sum_{k=1}^{n}\frac{\sin(\delta_s+k\pi/2)}{k!}(\delta-\delta_s)^k=0 \quad (12)$$

Recall the fact that $\delta_{u1}$ linearly depends on $\delta_s$ as $\delta_{u1} = \pi-\delta_s$. When $\delta_s$ ranges from 0 to $\pi/2$ with a small step of 0.001 rad $\approx$ 0.06°, numerically searching for $\delta_{u1\_TEn}$ by (12) for cases with $n=2, 3, …, 9$. Results are shown in Fig. 1 and Fig. 2.

Fig. 1 shows that the approximated UEPs $\delta_{u1\_TEn}$ are always smaller than $\delta_{u1}$ for cases with $n=3, 4, 7$ and 8, such that the stability analysis based on these TTE systems always give conservative results. The conservativeness decreases with the order, i.e. $\delta_{u1\_TE3} \leq \delta_{u1\_TE4} \leq \delta_{u1\_TE7} \leq \delta_{u1\_TE8} \leq \delta_{u1}$.

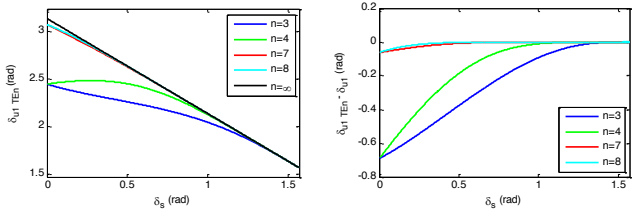

Fig. 1. $\delta_{u1\_TEn}$ and their errors for cases with $n$=3, 4, 7 and 8.

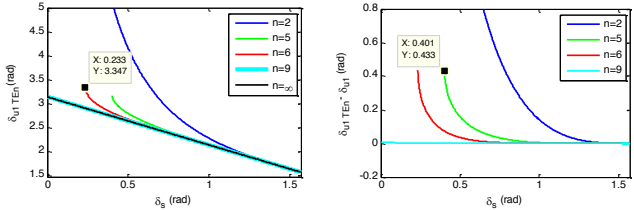

Fig. 2. $\delta_{u1\_TEn}$ and their errors for cases with $n$=2, 5, 6 and 9.

Fig. 2 shows that the approximated UEPs $\delta_{u1\_TEn}$, if existing, are always greater than $\delta_{u1}$ for cases with $n$=2, 5, 6 and 9, such that the stability analysis based on these TTE systems always gives optimistic results. The degree of optimism decreases with the order, i.e. $\delta_{u1\_TE2} \geq \delta_{u1\_TE5} \geq \delta_{u1\_TE6} \geq \delta_{u1\_TE9} \geq \delta_{u1}$. It should be noted that in the 5$^{th}$ order TTE system, when $\delta_s$ is smaller than 0.401 rad ≈ 22.98°, $\delta_{u1\_TE5}$ does not exist. Such phenomenon can be explained by the change in the sign of the discriminant $\Delta$ at $\delta_s \approx 0.401$ rad of the underlying quartic equation [10], as shown in Fig. 3. A similar phenomenon can also be observed in the 6$^{th}$ order TTE system which happens at $\delta_s \approx 0.233$ rad ≈ 13.35°. This can also be explained by discriminant [10]. In these two cases, when $\delta_{u1}$ disappears for very small $\delta_s$, i.e. the stability analysis using 5$^{th}$ or 6$^{th}$ order TTE systems always gives stable results for very unstressed loading conditions.

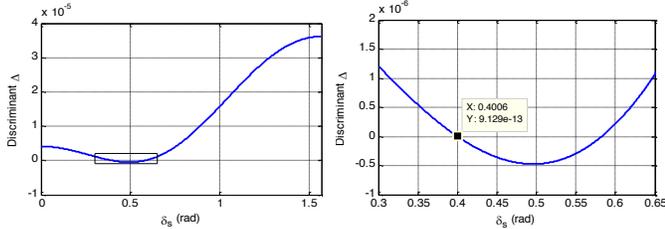

Fig. 3. The discriminant $\Delta(\delta_s)$ of the quartic equation for determining $\delta_{u1\_TE5}$.

### D. Transient stability analysis by power-angle curves

The relationship between the electrical power $P_e$ and the angle is described as the power-angle ($P$-$\delta$) curve, as shown in (13), which is often used to interpret the dynamics and stability of an SMIB power system (or the two-machine equivalent of multi-machine power system). Fig. 4 shows a typical $P$-$\delta$ curve and its approximations by $n^{th}$ order TTE systems, denoted as $P_{en}$. The right plot magnifies the black box in the left plot, from which $\delta_{u1\_TEn}$, i.e. the right intersection between $P_{en}$ and $P_m$, in 3$^{rd}$, 4$^{th}$, 7$^{th}$ and 8$^{th}$ order TTE systems are on the left side of $\delta_{u1}$, while on the right side in 2$^{nd}$, 5$^{th}$, 6$^{th}$ and 9$^{th}$ order TTE systems.

$$P_e = P_{max} \sin\delta \quad (13)$$

Then, the disappearance of $\delta_{u1\_TE5}$ and $\delta_{u1\_TE6}$ at small $\delta_s$ are verified by observing the approximated $P$-$\delta$ curves with $\delta_s$=20° and $\delta_s$=10°, which are shown in Fig. 4. When $\delta_s$ decreases from 30° to 20°, as shown in the left plot of Fig. 4 and the left plot of Fig. 5, the right intersection between $P_{e5}$ and $P_m$ disappears, i.e. $\delta_{u1\_TE5}$ disappears. Similarly, $\delta_{u1\_TE6}$ disappears when $\delta_s$ decreases from 20° to 10°. These observations are the same as the expectation from the analysis in subsection C.

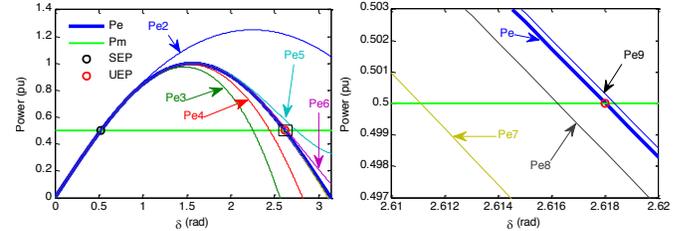

Fig. 4. A typical $P$-$\delta$ curve and its approximations by TTE systems with $P_{max}$ = 1 pu, $\delta_s$ = 30°.

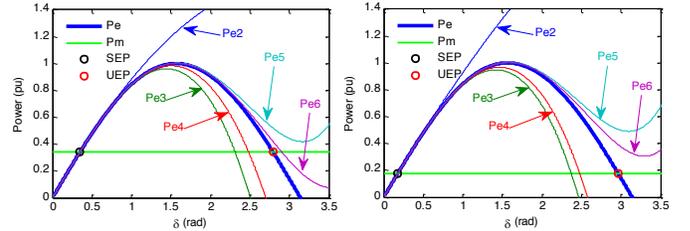

Fig. 5. $P$-$\delta$ curve and its approximations by TTE systems with $\delta_s$ = 20° (left) or $\delta_s$ = 10° (right).

Finally, Fig. 6 shows a case with a large steady-state angle, i.e. $\delta_s$ = 60°. It can be seen that when the system is operated close to its steady-state angle stability limit, $\delta_{u1\_TE2}$, …, $\delta_{u1\_TE5}$ are all fairly accurate estimates of $\delta_{u1}$. Note that $\delta_{u1\_TE6}$, …, $\delta_{u1\_TE9}$ almost overlap with $\delta_{u1}$.

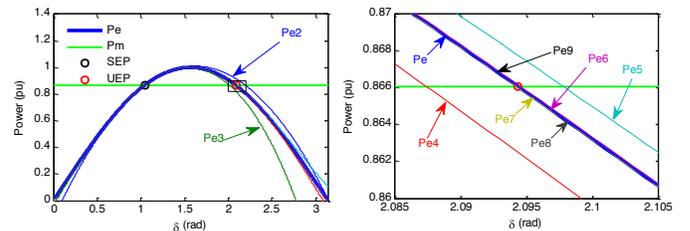

Fig. 6. $P$-$\delta$ curve and its approximations by TTE systems with $\delta_s$ = 60°.

### E. Remarks

This section investigates the transient stability analysis of the SMIB power system (1) using its TTE systems at different orders (3). The observations are summarized below:

- With a larger $n$, $\delta_{u1\_TEn}$ is closer to $\delta_{u1}$.
- $\delta_{u1\_TE5}$ does not exist when $\delta_s \leq 0.401$ rad.
- $\delta_{u1\_TE6}$ does not exist when $\delta_s \leq 0.233$ rad.
- When $\delta_{u1\_TEn}$ exist, these inequalities hold: $\delta_{u1\_TE2} \geq \delta_{u1\_TE5} \geq \delta_{u1\_TE6} \geq \delta_{u1\_TE9} \geq \delta_{u1}$. Thus, stability analysis results from TTE systems at these orders are optimistic.
- These inequalities always hold: $\delta_{u1\_TE3} \leq \delta_{u1\_TE4} \leq \delta_{u1\_TE7}$

$\leq \delta_{u1\_TE8} \leq \delta_{u1}$. So TTE systems at these orders always give conservative stability analysis results.

- If $\delta_s$ is small, i.e. the system is lightly loaded, then 3$^{rd}$ or 4$^{th}$ order TTE systems are recommended to perform a conservative stability analysis. If the computation burden is not a concern, then 7$^{th}$ or 8$^{th}$ order (or even higher orders) TTE systems are preferred.
- When $\delta_s$ is large, i.e. the system is heavily loaded, all $\delta_{u1\_TEn}$ for n = 2, 3, …, are good estimates of $\delta_{u1}$.
- A conjecture is made without a proof: **For SMIB power systems, their TTE systems at order $4n$ or $4n$-1 give conservative stability analyses, while TTE systems at order $4n$-2 or $4n$-3 give optimistic analyses ($n \geq 1$)**.
- Analytical approaches starting from a low order TTE system, e.g. 2$^{nd}$ order [6][7] or 3$^{rd}$ order [8][11], may be accurate for larger $\delta_s$ while inaccurate for small $\delta_s$.

## III. Transient Stability Analysis of Multi-Machine Power Systems Using TTE Systems

This section investigates the transient stability analysis of multi-machine power systems using TTE systems. In order to explore the capability of analytical approaches starting from TTE systems and without involving systematic errors, the transient stability is assessed by extensive time domain simulations in this section. A stability boundary searching algorithm is proposed to determine the stability boundaries of the original system and its TTE systems, whose comparison can show how accurately the TTE systems can be used to assess the transient stability of multi-machine power systems.

Consider an $m$-machine power system in classical model [12], whose ordinary differential equations are shown in (14). For each generator, the damping constant takes the same value as its inertia constant, such that the entire system has enough damping.

$$\ddot{\delta}_i + \frac{D_i}{2H_i}\dot{\delta}_i + \frac{\omega_s}{2H_i}(P_{mi} - P_{ei}) = 0 \tag{14a}$$

$$P_{ei} = E_i^2 G_i + \sum_{j=1, j \neq i}^{m} \left( C_{ij} \sin(\delta_i - \delta_j) + D_{ij} \cos(\delta_i - \delta_j) \right) \tag{14b}$$

### A. Stability boundary search algorithm

The stability boundaries of (14) and its TEE systems are identified by the algorithm below. Different parts of the stability boundary are acquired by applying sustained disturbances respectively in different directions in the state space. In this paper, the random number generation function "rand" in Matlab is used to create a number of unit vectors to represents disturbances in different directions in state space.

$$\mathbf{x} = \begin{pmatrix} \delta_1 & \Delta\omega_1 & \delta_2 & \Delta\omega_2 & \cdots & \delta_m & \Delta\omega_m \end{pmatrix}^T \tag{15}$$

### B. Tests on IEEE 9-bus power system

The one-line diagram and system parameters can be found in [12]. The number of all ($N$-1) line tripping contingencies is 12, which are shown in Table. II. The critical clearing time (CCT) of the original system is identified by a number of simulation runs and shown in Table II, while the CCTs of the TTE systems are normalized values by their corresponding reference CCTs from Table. II are presented in Table. III.

**START**
1. Let $k = 1$ and $l = l_0$, $s = s_0$.
2. Given a unit vector $\mathbf{n}$ in the state space representing the disturbance in a specific direction.
3. Initialize the system at $\mathbf{x}(t=0) = \mathbf{x}_{ep} + l \cdot \mathbf{n}$ and simulate the system for a certain time $T$.
4. **if** $|\delta_i(t=T) - \delta_j(t=T)| < \Delta$ for any $i$ and $j$ in $\{1, 2, …, m\}$
   | **if** $s < \varepsilon$
   | | $(\mathbf{x}_{ep} + l_k \cdot \mathbf{n})$ is on the boundary. **Return**.
   | **else**
   | | $l = l + s$.
   | **end**
   **else**
   | $s = s/2$ and $l = l - s$.
   **end**
5. **go to 3**

TABLE II  (N-1) LINE TRIPPING CONTINGENCIES AND THEIR CCTS

| Cont. # | Fault bus | Tripped line | CCT /s | Cont. # | Fault bus | Tripped line | CCT /s |
|---|---|---|---|---|---|---|---|
| 1 | 4 | 4-6 | 0.329 | 7 | 7 | 5-7 | 0.179 |
| 2 | 4 | 4-5 | 0.338 | 8 | 7 | 7-8 | 0.195 |
| 3 | 5 | 4-5 | 0.441 | 9 | 8 | 7-8 | 0.297 |
| 4 | 5 | 5-7 | 0.353 | 10 | 8 | 8-9 | 0.325 |
| 5 | 6 | 4-6 | 0.493 | 11 | 9 | 6-9 | 0.231 |
| 6 | 6 | 6-9 | 0.430 | 12 | 9 | 8-9 | 0.249 |

TABLE III  NORMALIZED CCTS OF TTE SYSTEMS IN 9-BUS SYSTEM

| # | TTE2 | TTE3 | TTE4 | TTE5 | TTE6 | TTE7 | TTE8 | TTE9 |
|---|---|---|---|---|---|---|---|---|
| 1 | 1.817 | 0.895 | 0.935 | 2.446 | 1.016 | 0.998 | 0.999 | 1.000 |
| 2 | 1.806 | 0.885 | 0.929 | 2.351 | 1.287 | 0.997 | 0.999 | 1.000 |
| 3 | 1.717 | 0.857 | 0.911 | ∞* | ∞ | 0.996 | 0.998 | 1.000 |
| 4 | 1.282 | 0.904 | 0.973 | 1.014 | 1.002 | 0.999 | 1.000 | 1.000 |
| 5 | 1.623 | 0.859 | 0.910 | ∞ | 1.024 | 0.996 | 0.998 | 1.000 |
| 6 | 1.373 | 0.892 | 0.960 | 1.020 | 1.004 | 0.999 | 1.000 | 1.000 |
| 7 | 1.276 | 0.908 | 0.974 | 1.013 | 1.002 | 0.999 | 1.000 | 1.000 |
| 8 | 1.347 | 0.911 | 0.969 | 1.015 | 1.003 | 0.999 | 1.000 | 1.000 |
| 9 | 1.380 | 0.895 | 0.959 | 1.023 | 1.005 | 0.999 | 1.000 | 1.000 |
| 10 | 1.602 | 0.870 | 0.929 | 2.361 | 1.016 | 0.997 | 0.999 | 1.000 |
| 11 | 1.358 | 0.904 | 0.964 | 1.019 | 1.004 | 0.999 | 1.000 | 1.000 |
| 12 | 1.776 | 0.887 | 0.930 | 2.345 | 1.061 | 0.997 | 0.999 | 1.000 |

* In each of these cases, the CCT is larger than 1 second.

Table III shows that the CCTs determined by TTE systems at orders 2, 5, 6 and 9 are larger than the true CCT, indicated by a number greater than one, while the CCTs from TTE systems at orders 3, 4, 7 and 8 are always smaller, indicated by a number smaller than one. In addition, the degree of optimism/conservativeness, i.e. the error in CCT, decreases when the TE order increases.

The second test is to consider a more stressed loading condition. The generations are re-dispatched in this way: increase $P_{m2}$ from 163MW to 200MW, $P_{m3}$ from 85MW to 100MW and decrease $P_{m1}$ from 71.61MW to 22.55MW. Such modification will push the system closer to its steady state angle stability limit due to the increase of the power transfer from generators 2 and 3 to generator 1. It should be mentioned that after the re-dispatch, the CCT of each contingency is reduced compared to their counterparts in Table. II, which is not shown here due to limited space. In this case, the normalized CCTs of TTE systems are shown in Table. IV. The comparison between Table. IV and Table. III indicates that the errors in the CCTs of TTE systems decrease if the system is closer to its steady state angle stability limit as expected from the observation on SMIB systems.

TABLE IV. NORMALIZED CCTs IF TTE SYSTEMS IN 9-BUS SYSTEM AFTER GENERATION RE-DISPATCH

| # | TTE2 | TTE3 | TTE4 | TTE5 | TTE6 | TTE7 | TTE8 | TTE9 |
|---|------|------|------|------|------|------|------|------|
| 1 | 1.409 | 0.907 | 0.964 | 1.018 | 1.004 | 0.999 | 1.000 | 1.000 |
| 2 | 1.344 | 0.897 | 0.963 | 1.024 | 1.007 | 0.999 | 1.000 | 1.000 |
| 3 | 1.388 | 0.887 | 0.959 | 1.025 | 1.005 | 0.999 | 1.000 | 1.000 |
| 4 | 1.894 | 0.592 | 0.941 | 1.030 | 1.001 | 0.999 | 0.999 | 0.999 |
| 5 | 1.430 | 0.894 | 0.958 | 1.021 | 1.005 | 0.999 | 1.000 | 1.000 |
| 6 | 1.173 | 0.929 | 0.989 | 1.004 | 1.000 | 1.000 | 1.000 | 1.000 |
| 7 | 1.854 | 0.577 | 0.942 | 1.027 | 0.999 | 0.997 | 0.998 | 0.998 |
| 8 | 1.206 | 0.923 | 0.985 | 1.006 | 1.001 | 1.000 | 1.000 | 1.000 |
| 9 | 1.222 | 0.919 | 0.983 | 1.007 | 1.001 | 1.000 | 1.000 | 1.000 |
| 10 | 1.387 | 0.894 | 0.962 | 1.020 | 1.004 | 0.999 | 1.000 | 1.000 |
| 11 | 1.176 | 0.926 | 0.989 | 1.005 | 1.001 | 1.000 | 1.000 | 1.000 |
| 12 | 1.521 | 0.887 | 0.943 | 1.151 | 1.012 | 0.998 | 0.999 | 1.000 |

The last test investigates the TTE based transient stability analysis subject to 1000 randomly generated directions in the state space under the default loading condition to represent other disturbances not covered by the 12 contingencies in Table. II. The distributions of the normalized stability boundaries of TTE systems are shown in Fig. 7, which verifies the observations on SMIB power systems. Specifically, $2^{nd}$, $6^{th}$ and $9^{th}$ order TTE systems generally give the stability boundaries respectively 0.7-2.5, 0.66-3, 1-1.005 times larger compared to the true boundaries, while the $5^{th}$ order is about 1-5 times larger including a few cases where an instability cannot be detected. The stability boundaries from $3^{rd}$, $4^{th}$, $7^{th}$ and $8^{th}$ order TTE systems are respectively about 0.52-0.88, 0.51-0.96, 0.98-0.999 and 0.99-1 times the true boundaries.

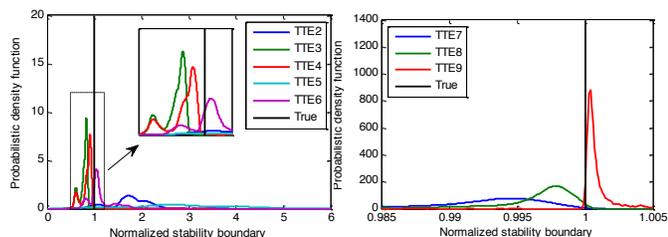

Fig. 7. Distributions of normalized stability boundaries of TTE systems of IEEE 9-bus power system in 1000 different directions in the state space.

### C. Tests on WECC 179-bus power system

A simplified WECC power system model with 179 buses and 29 generators is used to test the accuracy of TTEs on top-8 most critical line-tripping contingencies near the California-Oregon intertie [13]. The results shown in Table V also validate the observations on SMIB systems. Recall the comparison between Table III and Table IV, the accurate estimates of stability boundaries in Table V from TTE systems at order 4 order and above indicate that the original system is close to its steady state angle stability limit along the California-Oregon intertie.

## IV. CONCLUSIONS

This paper explores the possibility of using truncated Taylor expansion (TTE) systems to study the transient stability of a power system. Analytical investigations on SMIB systems observe that TTE systems at orders 2, 5, 6 and 9 give optimistic transient stability analysis results while those at orders 3, 4, 7 and 8 give conservative results. Numerical studies on two multi-machine power systems show that such phenomena are also observed in multi-machine power systems. These observations lay a foundation for all analytical stability analysis approaches starting from TTE systems.

TABLE V. NORMALIZED CCTs OF TTE SYSTEMS IN 179-BUS SYSTEM

| # | TTE2 | TTE3 | TTE4 | TTE5 | TTE6 | TTE7 | TTE8 | TTE9 |
|---|------|------|------|------|------|------|------|------|
| 1 | 1.144 | 0.885 | 0.996 | 1.017 | 1.000 | 0.999 | 1.000 | 1.000 |
| 2 | 1.144 | 0.886 | 0.995 | 1.017 | 1.000 | 0.999 | 1.000 | 1.000 |
| 3 | 1.104 | 0.895 | 1.000 | 1.014 | 1.000 | 0.999 | 1.000 | 1.000 |
| 4 | 1.105 | 0.894 | 1.000 | 1.015 | 1.000 | 0.999 | 1.000 | 1.000 |
| 5 | 1.085 | 0.917 | 0.996 | 1.009 | 1.000 | 1.000 | 1.000 | 0.999 |
| 6 | 1.086 | 0.916 | 0.996 | 1.009 | 1.000 | 1.000 | 1.000 | 1.000 |
| 7 | 1.116 | 0.893 | 1.040 | 1.015 | 1.000 | 0.999 | 1.000 | 1.000 |
| 8 | 1.119 | 0.893 | 1.041 | 1.015 | 1.000 | 0.999 | 1.000 | 1.000 |